\documentclass[
 prb, superscriptaddress,
 amssymb,
 preprint,
 11pt
]{revtex4-2}

\usepackage{dcolumn}
\usepackage{bm}
\usepackage[paperwidth=21.6cm,paperheight=28cm,centering,hmargin=2.54cm,vmargin=2.54cm]{geometry}
\usepackage{xcolor,soul,framed} 
\colorlet{shadecolor}{yellow}
\usepackage[pdftex]{graphicx}
\graphicspath{{../pdf/}{../jpeg/}}
\DeclareGraphicsExtensions{.pdf,.jpeg,.png}
\usepackage[colorlinks,citecolor=blue,urlcolor=blue,linkcolor=blue,bookmarks=false,hypertexnames=true]{hyperref}
\usepackage[cmex10]{amsmath}
\usepackage{cleveref}
\Crefformat{figure}{#2Figure~#1#3}
\usepackage{array}
\usepackage{mdwmath}
\usepackage{mdwtab}
\usepackage{eqparbox}
\usepackage{url}
\usepackage{bibentry}
\usepackage[justification=raggedright,singlelinecheck=false,labelfont=bf,labelsep=quad, format=plain,font=small]{caption,subcaption}
\usepackage{orcidlink}

\begin{document}
\title{Investigation of Cryogenic Current-Voltage Anomalies in SiGe HBTs:\\Role of Base-Emitter Junction Inhomogeneities}

\author{Nachiket R. Naik\,\orcidlink{0000-0001-8271-643X}}
\affiliation{ Division of Engineering and Applied Science, California Institute of Technology, Pasadena, California, 91125, USA}
\author{Bekari Gabritchidze\,\orcidlink{0000-0001-6392-0523}}
\affiliation{ Cahill Radio Astronomy Lab, California Institute of Technology, Pasadena, California, 91125, USA}
\author{Justin H. Chen\,\orcidlink{0000-0001-9745-7055}}
\affiliation{ Division of Engineering and Applied Science, California Institute of Technology, Pasadena, California, 91125, USA}
\author{Kieran A. Cleary\,\orcidlink{0000-0002-8214-8265}}
\affiliation{ Cahill Radio Astronomy Lab, California Institute of Technology, Pasadena, California, 91125, USA}
\author{Jacob Kooi\,\orcidlink{0000-0002-6610-0384}}
\affiliation{ NASA Jet Propulsion Laboratory, California Institute of Technology, Pasadena, California, 91109, USA}
\author{Austin J. Minnich\,\orcidlink{0000-0002-9671-9540}}
 \email{aminnich@caltech.edu.}
\affiliation{ Division of Engineering and Applied Science, California Institute of Technology, Pasadena, California, 91125, USA}

\date{\today}


\begin{abstract}
The deviations of cryogenic collector current-voltage characteristics of SiGe heterojunction bipolar transistors (HBTs) from ideal drift-diffusion theory have been a topic of investigation for many years. Recent work indicates that direct tunneling across the base contributes to the non-ideal current in highly-scaled devices. However, cryogenic discrepancies have been observed even in older-generation devices for which direct tunneling is negligible, suggesting another mechanism may also contribute. Although similar non-ideal current-voltage characteristics have been observed in Schottky junctions and were attributed to spatial inhomogeneities in the base-emitter junction potential, this explanation has not been considered for SiGe HBTs. Here, we experimentally investigate this hypothesis by characterizing the collector current ideality factor and built-in potential  of a SiGe HBT versus temperature using a cryogenic probe station. The temperature-dependence of the ideality factor and the relation between the built-in potential as measured by capacitance-voltage and current-voltage characteristics are in good qualitative agreement with the predictions of a theory of electrical transport across a junction with a Gaussian distribution of  potential barrier heights. These observations suggest that lateral inhomogeneities in the base-emitter junction potential may contribute to the cryogenic non-idealities. This work helps to identify the physical mechanisms limiting the cryogenic microwave noise performance of SiGe HBTs.
\end{abstract}

\keywords{Heterojunction bipolar transistors, SiGe, cryogenic microwave amplifiers, device physics, barrier inhomogeneities, ideality factor, microwave noise.}

\maketitle
\newpage

\section{\label{sec:intro}Introduction}
Silicon-germanium heterojunction bipolar transistors (HBTs) are widely used in  microwave applications such as high-speed communications and radar systems owing to their competitive microwave performance, low cost, and ease of integration compared with III-V compound semiconductor devices \cite{cressler2003silicon}. Technological advances such as reduced emitter widths, decreased base resistances and extrinsic capacitances, and advanced epitaxial techniques have enabled microwave noise performance approaching that of III-V high electron mobility transistors \cite{Chevalier2017, Wong2020a}. As a result, cryogenic SiGe HBTs have recently been considered for applications in  quantum computing and radio astronomy \cite{Bardin2017a, Ying2018, bardin_2021a}.

The cryogenic microwave performance of  SiGe HBTs has been investigated following their initial development in the 1980s \cite{Patton1988a} as various performance metrics such as transconductance and noise figure improve with cooling. However, below $\sim$ 77 K, these improvements are observed to plateau with decreasing temperature \cite{Joseph1995, richey1996, Bardin2009a}, corresponding to a temperature-dependent collector current ideality factor $n(T)$ that greatly exceeds unity at cryogenic temperatures \cite{Rucker2017, Jin2021}. This behavior differs markedly from the predictions of drift-diffusion theory for fabricated p-n junctions for which $n$ is close to unity and independent of temperature, and the transconductance increases inversely with temperature. \cite{Monch2001, sze2021}  Because the minimum noise temperature is directly proportional to $n(T)$ in the low-frequency and low base-resistance limit (see \cite[Eq.~2]{Bardin2017a}), identifying the physical origin of the discrepancies is necessary to improve the cryogenic microwave noise performance of SiGe HBTs.

The cryogenic non-ideal behavior has been attributed to various mechanisms including quasiballistic transport \cite{Bardin2009a, richey1996}, direct tunneling \cite{Rucker2017}, or trap-assisted tunneling \cite{Davidovic2017}. However, a theoretical study has reported  that quasiballistic electron transport cannot explain the observed collector cryogenic non-idealities \cite{Naik2021}. Recent works indicate that direct tunneling across the base can account for the non-idealities in highly-scaled devices \cite{Jin2021, Rucker2017, Ying2018, schroter_2023}. At the same time,  non-idealities have been observed in devices with base widths of $\sim$100 nm \cite{richey1996, Ying2018} for which direct tunneling is negligible. This observation suggests that, in addition to tunneling, another mechanism may contribute to cryogenic non-ideal current-voltage characteristics in SiGe HBTs.

In a different context, similar anomalies have been observed and extensively investigated in Schottky diodes \cite{Hackam1972, padovani_1965, saxena_1969, ashok_1979}, and they were ultimately attributed to spatial inhomogeneities in the built-in potential $\Phi_{BI}$. \cite{Tung1992, werner1991barrier, freeouf_1982, song_1986}  Although semiconductor junctions are often modeled as uniform across the lateral area, in fact various imperfections exist which affect the local electronic structure of the junction, a point which was recognized as early as 1950. \cite{johnson_1950}  Even at epitaxial interfaces,  it was found that different crystallographic orientations \cite{tung_1984, hauenstein_1985, tung_1991a} or the presence of dislocations \cite{knel_1998, vonknel_1997} can lead to potential barrier height variations on the order of hundreds of mV. In Schottky junctions, these inhomogeneities have been directly observed using ballistic electron emission microscopy. \cite{kaiser_1988, prietsch1995}  Various theories and numerical analyses of the electrical characteristics of inhomogeneous junctions have been reported and lead to compatible conclusions. \cite{freeouf_1982, Tung1992, werner1991barrier} For concreteness, the theory of Werner and G{\"u}ttler assumes a Gaussian distribution of barrier heights and makes several  predictions regarding the temperature-dependence of the ideality factor and the relation between the built-in potential as measured by different methods. \cite{werner1991barrier, werner_1991_physica_scripta, rau_1992}  However, an experimental test of these predictions for SiGe HBTs has not yet been reported.


Here, we perform this experimental investigation by characterizing the collector current ideality factor and built-in potential of a SiGe HBT  from room to cryogenic temperatures. We find that the measured temperature-dependence of the ideality factor and the relation between built-in potential as determined by capacitance-voltage and current-voltage characteristics are compatible with the theory predictions. This observation suggests that inhomogeneities in the base-emitter junction potential could be a mechanism affecting the cryogenic current-voltage characteristics in SiGe HBTs.   We discuss how the existence of barrier inhomogeneities could be further confirmed. Our work advances efforts to improve the cryogenic electrical characteristics and hence microwave noise performance of SiGe HBTs.





\section{\label{sec:th&met} Theory and Methods}

\subsection{\label{sec:Overview}Overview}
The theory of Werner and G{\"u}ttler describes electrical transport across a potential barrier with heights distributed according to a Gaussian distribution of a specified mean and variance. \cite{werner1991barrier} The variance of the distribution is assumed to decrease with increasing junction bias due to the pinch-off of low-barrier patches of dimension less than the depletion length, a concept that was originally introduced in \cite{freeouf_1982} and later developed in \cite{Tung1992}. The theory makes several predictions regarding the trends of electrical characteristics with temperature and other parameters in junctions exhibiting voltage-independent ideality factors $n(T)$. In particular, $n(T)$, as determined from the slope of $I-V$ characteristics, is predicted to vary with temperature according to $n(T)^{-1} - 1 = -\rho_2 + \rho_3/(2 k T/q)$ where $k$ is Boltzmann's constant, $q$ is the electric charge, and $\rho_2$ and $\rho_3$ are constants describing the variation of the mean and variance of the potential barrier distribution with junction voltage, respectively (see \cite[Eq.~23]{werner1991barrier}). A plot of $n(T)^{-1} - 1$ versus $T^{-1}$ should therefore yield a line over some range of temperatures.

In addition, the built-in potential $\Phi_{BI}$ can be measured in two ways. From $C_{BE}-V_{BE}$ characteristics, $\Phi_{BI}(CV)$ can be obtained by fitting the  variation of depletion capacitance with junction voltage using $C_{BE}(V_{BE}) = C_{BE,0} (1 - V_{BE}/\Phi_{BI})^{-m}$, where $C_{BE,0}$ is the zero-bias junction capacitance and $m$ is an exponent that depends on the doping profile at the junction. \cite{sze2021} Using $I_C-V_{BE}$ characteristics, the potential barrier energy for transport, $E_a$, relative to its value at some temperature, can be obtained by extrapolating the measured collector current  to zero bias using the collector current expression $I_C  = I_S (\exp( {q V_{BE}/ n(T ) k T} ) - 1)$ where $I_S = A \exp( -q E_a/k T)$ is the transport saturation current and $A$ is a constant prefactor. \cite[Sec.~4.2.1]{cressler2003silicon} We have neglected the polynomial temperature-dependence of the prefactor as it only makes a log-scale correction to the built-in potential which does not alter our conclusions.

With the value of the transport barrier at 300 K specified as $\Phi_{BI}(CV)$, the temperature-dependence of the apparent built-in potential associated with charge transport $\Phi_{BI}(IV)$ can thus be obtained from the value of $I_S$ at each temperature relative to that at the reference temperature. We note that this charge injection approach to model the current-voltage characteristics is relatively simple compared to generalized-integral charge-control relation (GICCR) models employed for HBTs. \cite{pawlak_2014} However, GICCR models also exhibit discrepancies with experiment below 100 K, \cite[Fig.~6.5(a)]{jin_thesis} suggesting that the physical origin is not due to the simplicity of the charge injection model.

The inhomogeneous junction theory predicts that the barrier measured in these two ways should differ in magnitude and temperature dependence owing to the fact that current depends on $V_{BE}$ exponentially while the capacitance varies with $V_{BE}$ with a weaker polynomial dependence.  The barrier as determined through $C-V$ characteristics is therefore typically interpreted as the mean barrier height, while that determined from $I-V$ characteristics is often less than the mean value due to the larger contribution from low-barrier regions. \cite{ohdomari_1979, werner1991barrier} The theory gives a relation between these barrier heights. \cite[Eq.~14]{werner1991barrier} Considering the form of $n(T)$ described above, this relation is compatible with the empirical observation that $\Phi_{BI}(CV) \approx n(T) \Phi_{BI}(IV)$ \cite{Bhuiyan1988a} (see also \cite[Sec.~V]{werner1991barrier}). 

These predictions can be tested using a cryogenic probe station to measure the $I_C-V_{BE}$ and $C_{BE}-V_{BE}$ characteristics of a SiGe HBT. We note that for the $C_{BE}-V_{BE}$ characteristics, although the mechanisms of current transport in the forward active regime differ between Schottky junctions and HBTs, the electrostatics of the depletion capacitance are identical between the devices. \cite[Chaps.~6 and 7]{delalamo_2017} Therefore, the built-in potential can be determined by the dependence of $C_{BE}$ on $V_{BE}$  for biases for which $C_{BE}$ is dominated by the depletion capacitance. This procedure was applied to determine the built-in potential of SiGe HBTs in another study. \cite{Jin2021}



\subsection{Experimental methods} \label{sec:experimental-methods}

We extracted $\Phi_{BI}$ from $C_{BE}-V_{BE}$ and $I_C - V_{BE}$ characteristics from 20 -- 300 K on a SiGe HBT (SG13G2, IHP). The discrete transistors were probed in a custom-built cryogenic probe station. \cite{Gabritchidze2022, russell2012cryogenic} We employed Nickel/Tungsten probes (40A-GSG-100-DP, GGB Industries) which are suitable for probing Al pads. $I_C - V_{BE}$ characteristics were performed at a constant collector voltage $V_{CE} =1$ V to provide a collector current above the minimum resolution of our  measurement setup (10 nA). The current range used for this fitting is limited to 0.2 mA, below the high-injection regime, to exclude effects of series resistance, self-heating and the Early and reverse Early effects. Inclusion of the reverse Early effect in the extraction of the built-in potential was found to alter the value by only a few percent as it enters the expression for the $\Phi_{BI}$ as a log-scale correction. $\Phi_{BI}$ was extracted relative to its value determined by $C_{BE}-V_{BE}$ measurements at 300 K from the zero-bias intercept of a linear fit of $\ln(I_C)$ versus $V_{BE}$, and $n(T)$ was extracted from the slope of this fit. Due to the difficulty of distinguishing periphery  from area currents, we assumed that the area current is dominant following other studies of cryogenic SiGe HBTs. \cite{Rucker2017, schroter_2023, Jin2021, Ying2018, Bardin2009a}


Following standard procedure, \cite{Jin2021, Monch2001} $C_{BE}-V_{BE}$  characteristics were obtained using a vector network analyzer (VNA, Keysight E5061B). In reverse-bias and low-forward bias regimes, the $Y$-parameters are given by $Y_{11} = g_{BE} + j\omega( C_{BE} +C_{BC})$ and $Y_{12} = -j\omega C_{BC}$, where $C_{BC}$ is the base-collector depletion capacitance.  The base-emitter capacitance can therefore be expressed as $C_{BE} = (\Im(Y_{11} + Y_{12}))/2 \pi f$. $V_{BE}$ was restricted to [$-0.5$ V, $+0.5$ V] to ensure that the measured capacitance was dominated by the depletion capacitance. $V_{BC} = $ 0 V was held constant to ensure $C_{BC}$ was constant while $V_{BE}$ was swept. The $Y$ parameters were measured in $1-3$ GHz, and the extractions were performed at 2.4 GHz. At these frequencies, it was observed that the imaginary part of $Y_{11}$ was linear in frequency, indicating purely capacitive behavior. Short-Open-Load-Through calibration was performed on a CS-5 calibration standard at each temperature, and the shunt parasitic capacitance at the input of the device was de-embedded using an OPEN structure. The intermediate-frequency bandwidth (1 kHz) and frequency points (every 0.2 GHz) were selected to limit the total sweep time to less than 15 s to avoid drift. At each bias, Y-parameters were swept across frequency and ensemble-averaged 10 times. $\Phi_{BI}$ was extracted from a sweep of $C_{BE}$ versus $V_{BE}$ by fitting the parameters $\Phi_{BI}$, $C_{BE,0}$ and $m$ using a trust region reflective algorithm from the SciPy library. \cite{2020SciPy-NMeth} $\Phi_{BI}$ was constrained to [0.5 V, 1.2 V], $C_{BE,0}$ was constrained between the minimum and maximum values of the sweep, and $m$ was constrained to [0, 1].


\section{\label{sec:results} Results}

\begin{figure}
\centering
{\includegraphics[width= 0.7\textwidth]{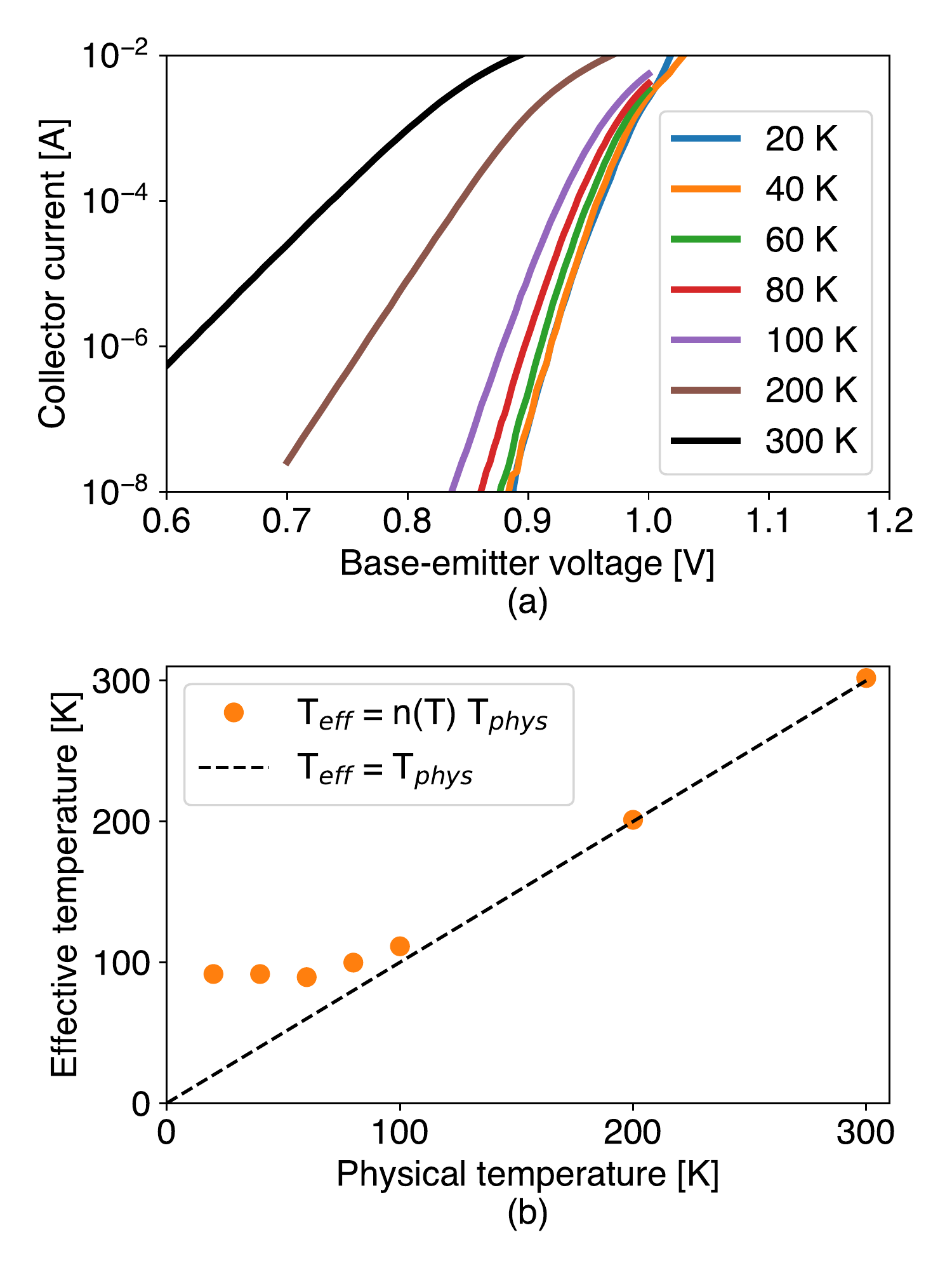}
    \phantomsubcaption\label{fig:Ic}
    \phantomsubcaption\label{fig:nTTeff}}

\caption{(a) Measured $I_C$ versus $V_{BE}$ for various temperatures. The characteristics become independent of temperature at cryogenic temperatures. (b) $T_{eff} = n(T) T_{phys}$ vs $T_{phys}$ from measurements (symbols) and diode theory (line), indicating the non-ideality of the base-emitter junction at cryogenic temperatures. }
\end{figure}

\begin{figure}
\centering%
{\includegraphics[width= 0.7\textwidth]{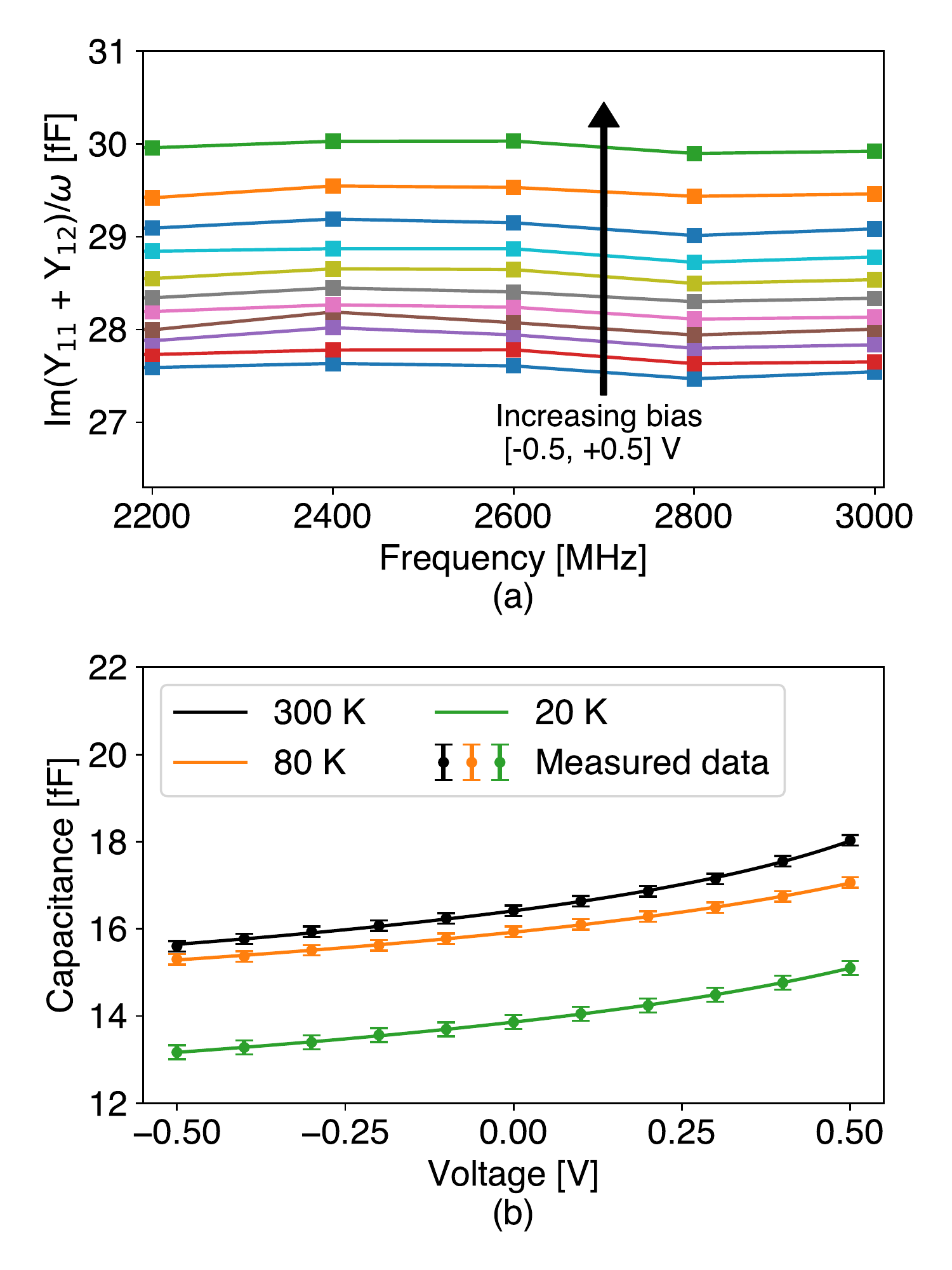}
    \phantomsubcaption\label{fig:Y11}
    \phantomsubcaption\label{fig:Cfit}}
    
\caption{(a) Measured $\Im(Y_{11} + Y_{12})/\omega$ (symbols) versus $f$  at 300 K for various $V_{BE}$ in steps of 0.1 V. The lines are guides to the eye.  (b) Measured (symbols) and fitted (solid lines) $C_{BE}$ versus $V_{BE}$ at 300 K, 80 K and 20 K. As a representative example, the fit for 300 K yields $\Phi_{BI} = 0.83$ V and $m = 0.10$.}
\end{figure}

\Cref{fig:Ic} shows the collector current $I_C$ versus $V_{BE}$ at various temperatures between 20 K and 300 K. Consistent with prior findings, \cite{Ying2018, Bardin2009a, Rucker2017} the measurements exhibit deviations with drift-diffusion theory at cryogenic temperatures, with the current-voltage characteristic plateauing to a temperature-independent curve below $\sim$60 K.  We plot the extracted $n(T)$ as $T_{eff} = n(T) T_{phys}$ versus $T_{phys}$ in \Cref{fig:nTTeff}. $T_{eff}$ is observed to  plateau to $\sim$100 K due to $n(T) > 1$ at cryogenic temperatures, as has been reported previously \cite{Bardin2009a}.

We next examine the RF characteristics. \Cref{fig:Y11} plots $\Im(Y_{11} + Y_{12})/\omega $ versus $f$ for various $V_{BE}$ at 300 K, where $\omega = 2 \pi f$. A narrowed frequency range from the $1-3$  GHz measurements is plotted to aid in distinguishing the curves. The de-embedded  base-emitter capacitance $C_{BE}$ is directly obtained from this plot by averaging across the frequency range. \Cref{fig:Cfit} plots the resulting $C_{BE}$ versus $V_{BE}$ at three representative temperatures across the overall voltage range along with the fitted curves. The error bars, representing the 2$\sigma$ error in $C_{BE}$, are obtained from the 10 $C_{BE}-V_{BE}$ sweeps performed at each temperature.

\begin{figure}[h]
\centering
{\includegraphics[width= 0.695\textwidth]
{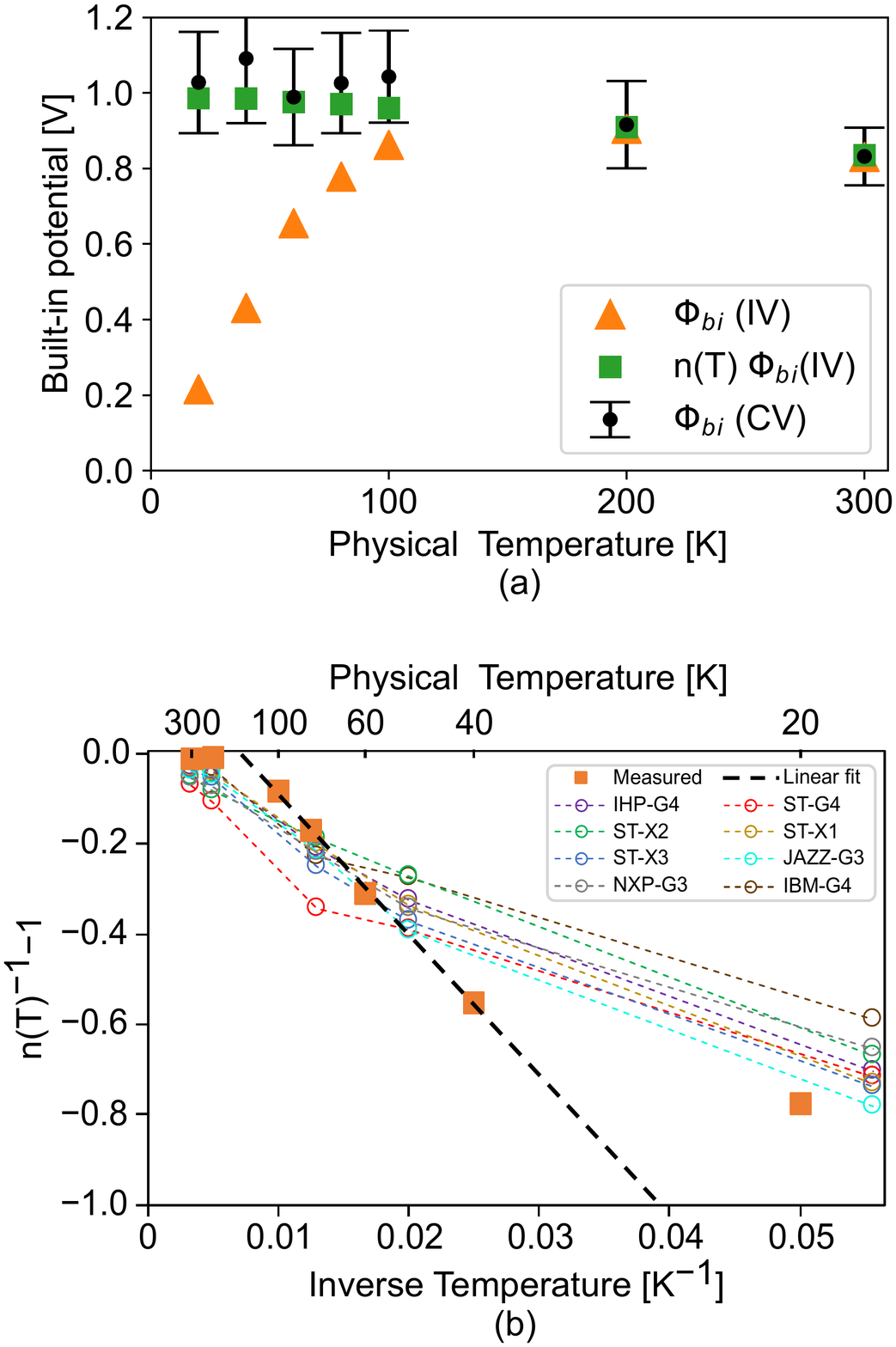}

    \phantomsubcaption\label{fig:Vb}
    \phantomsubcaption\label{fig:nTfit}}
\caption{(a) Built-in potential $\Phi_{BI}$ versus physical temperature from $C_{BE}-V_{BE}$ (black circles) and $I_C-V_{BE}$ (orange triangles) measurements. Also plotted is $n(T) \Phi_{BI} (IV)$ (green squares) which is predicted to agree with $\Phi_{BI}(CV)$. \cite{werner1991barrier} Good agreement is observed. (b) $n(T)^{-1} -1 $ versus inverse physical temperature $T^{-1}$ for measured data along with a linear fit to the data in 40-100 K following the prediction in \cite{werner1991barrier}. Data obtained from \cite[Fig.~5.8]{Bardin2009a} are also shown. }
\end{figure}

These data are next analyzed to obtain $\Phi_{BI}$ from the $I_C - V_{BE}$ and $C_{BE} - V_{BE}$ characteristics according to the procedures in \Cref{sec:experimental-methods}. At 300 K, $\Phi_{BI}(CV)$  is found to be 0.83 V, in good agreement with \cite{Jin2021}. This value is specified as the room temperature value for  $\Phi_{BI}(IV)$ to facilitate comparison at other temperatures. \Cref{fig:Vb} plots the $\Phi_{BI}$ from both measurements versus $T_{phys}$. For $\Phi_{BI}(CV)$, the error bars represent the 2-$\sigma$ error in $\Phi_{BI}$, obtained by performing fits to 100 $C_{BE} - V_{BE}$ sweeps with errors randomly determined based on a normal distribution defined by the uncertainty in the measured $C_{BE}$. The extracted $\Phi_{BI}(CV)$ is observed to weakly increase with decreasing temperature, consistent with observations for similar HBT devices \cite{Jin2021} and Schottky diodes \cite{Bhuiyan1988a}. In contrast, $\Phi_{BI}(IV)$ exhibits a qualitatively stronger dependence on temperature than  $\Phi_{BI}(CV)$ and differs markedly in magnitude at cryogenic temperatures, as observed previously for Schottky diodes \cite{Bhuiyan1988a, Monch2001} (also compare to \cite[Fig.~3]{werner_1991_physica_scripta}). The variation with temperature of  $\Phi_{BI}(IV)$ is significantly stronger than that of the emitter and base band gaps, \cite{sze2021} suggesting another mechanism is responsible for the observed temperature trend.

We now examine the agreement between the data and the predictions from the inhomogeneous junction theory of \cite{werner1991barrier}. First, qualitatively, $\Phi_{BI}(CV)$ and $\Phi_{BI}(IV)$ are predicted to differ, with $\Phi_{BI}(IV)$ expected to exhibit a stronger temperature dependence and be smaller in magnitude than $\Phi_{BI}(CV)$. This behavior is observed in \Cref{fig:Vb}. More quantitatively, for the predicted temperature-dependence of $n(T)$ from the theory, it is expected that $\Phi_{BI}(CV) \approx n(T) \Phi_{BI}(IV)$ \cite[Sec.~V]{werner1991barrier}. The product $n(T) \Phi_{BI}(IV)$ is also plotted in \Cref{fig:Vb}, demonstrating good agreement with $\Phi_{BI}(CV)$.

Second, \cite[Eq.~23]{werner1991barrier} predicts that  $n(T)^{-1} - 1$ versus $T^{-1}$ should be a straight line over some range of temperatures. \Cref{fig:nTfit} plots this quantity for the present device and other devices with data obtained from \cite[Fig.~5.8]{Bardin2009a}. For all the devices, the expected trend is observed over a temperature range which is comparable in relative width to that in \cite[Fig.~9]{werner1991barrier}. For the older-generation devices, the linear trend is observed from 18 -- 100 K. For the present device, a linear regime is found from 40 -- 100 K, while a deviation is noted at 20 K. These data are consistent with the absence of tunneling in the older devices and a non-negligible direct tunneling current in the present device. Other explanations for the deviation could include the assumption of temperature-independent barrier properties in the inhomogeneous barrier theory. However, overall the experimental trends are in good qualitative agreement with the theoretical predictions.



Semi-quantitative information regarding the variation of the barrier height variance with bias for the present device can be obtained from the linear fit in \Cref{fig:nTfit}. If $\rho_2 + \rho_3/(2 k T/q) \ll 1$, the theory of \cite{werner1991barrier} reduces to the $T_0$ model for non-ideal junctions which has been extensively studied in the Schottky junction literature. \cite{padovani_1965, Tung1992, werner1991barrier} In this case, the slope of the linear fit in \Cref{fig:nTfit} is simply $T_0$. Performing this fit for the present data yields $T_0 \approx 30$ K, a value which is generally compatible with values for Schottky diodes compiled in \cite{werner1991barrier}. Using \cite[Eq.~33]{werner1991barrier}, this value can be recast in terms of the narrowing of the potential barrier height distribution with increasing bias using $\rho_3 \approx - 2 k T_0/q$. We obtain a value $\rho_3 = - 5.2$ mV. The magnitude of this value indicates that changes in standard deviation of the potential barrier height distribution with base-emitter bias which are less than a percent of the mean barrier height are sufficient to account for the observed electrical anomalies.

\section{\label{sec:disc} Discussion}

The agreement of our data with the predictions of the inhomogeneous barrier theory suggests that lateral inhomogeneities in the base-emitter junction potential could contribute to the cryogenic electrical anomalies. Additional evidence for the barrier inhomogeneity hypothesis could be obtained using techniques such as ballistic emission electron microscopy which directly measures the spatial profile of the built-in potential. \cite{prietsch1995} However, applying this method to SiGe HBTs would require specialized samples to be prepared which are compatible with the measurement technique.



The materials-scale origin of the inhomogeneities in HBTs could be Ge clusters \cite{Kiehl1993} or electrically active carbon defects \cite{Raoult2008}. Non-uniform Ge content over a few nanometers in SiGe p-wells with Ge concentration $\geq$ 30\% has been reported to lead to the degradation of electrical properties like hole mobility. \cite{Kiehl1993} Trap states associated with C impurities have also been detected in modern HBTs. \cite{Raoult2008} With Ge concentrations for modern HBTs being on the order of 30\% \cite{Rucker2017} and C doping on the order of $10^{20}$ cm\textsuperscript{-3}, \cite{cressler2003silicon} these defects lead to an inhomogeneous base-emitter junction potential. If the presence of spatial inhomogeneities across the emitter area is verified, a less aggressive Ge doping concentration and profile, especially in narrow-base SiGe HBTs, could decrease the concentration of these imperfections and thereby lead to a more uniform base-emitter junction potential. However, care would need to be taken to avoid negatively impacting the high-frequency properties of the device. Additionally, direct tunneling may pose a fundamental obstacle to improving the cryogenic electrical ideality of highly-scaled devices. Further study is needed to distinguish the various mechanisms and identify the  limits to cryogenic electrical ideality and hence microwave noise performance.



\section{\label{sec:concl} Conclusion}
We have reported a characterization of the collector-current ideality factor and built-in potential versus temperature of a SiGe HBT. The observed trends with temperature and between measurement techniques agree with a theory of electrical transport at a potential barrier with a Gaussian distribution of barrier heights, suggesting that barrier inhomogeneities may contribute to cryogenic electrical non-idealities in SiGe HBTs. This work advances efforts to improve the cryogenic microwave noise performance of  SiGe HBTs.

\section*{Acknowledgments}
The authors thank Akim Babenko, John Cressler, Nicolas Derrier, Xiaodi Jin, Pekka Kangaslahti, Holger R{\"u}cker, Michael Schr{\"o}ter,  and Sander Weinreb for useful discussions. This work was supported by NSF Award Number 1911926 and by JPL PDRDF Project Number 107978.

\section*{Author Declarations}
\subsection*{Conflict of Interest} The authors have no conflicts to disclose.

\section*{Data Availability}
The data that support the findings of this study are available from the corresponding author upon reasonable request.

\bibliography{refs}

\end{document}